\documentclass[aps, prl, twocolumn, superscriptaddress]{revtex4-2}

\usepackage{hyperref, graphicx, amsmath, amssymb, siunitx, bm, upgreek}
\usepackage[dvipsnames]{xcolor}
\usepackage[caption=false, subrefformat=parens, labelformat=parens]{subfig}
\usepackage{mathtools}
\usepackage{braket}
\usepackage[normalem]{ulem}

\hypersetup{colorlinks, linkcolor=teal, citecolor=blue, urlcolor=purple}

\newcommand{\ba}{\hat{a}}
\newcommand{\bphi}{\hat{\phi}}

\newcommand{\bp}{\hat{\phi}}
\newcommand{\bvarp}{\hat{\varphi}}
\newcommand{\bH}{\hat{\mathcal{H}}}
\newcommand{\bD}{\hat{D}}
\newcommand{\thetaext}{\theta_\text{ext}}
\newcommand{\phiext}{\varphi_\text{ext}}
\newcommand{\phizpf}{\eta}
\newcommand{\EJ}{\mathcal{E}_J}
\newcommand{\varphizpf}{\varphi_\text{zpf}}

\newcommand{\LPENS}{\affiliation{Laboratoire de Physique de l'Ecole Normale Supérieure, ENS-PSL, CNRS, Sorbonne Université, Université Paris Cité, Centre Automatique et Systèmes, Mines Paris, Université PSL, Inria, Paris, France}}

\newcommand{\LPTHE}{\affiliation{Laboratoire de Physique Th\'{e}orique et Hautes Energies, Sorbonne Universit\'{e} and CNRS UMR 7589, 4 place Jussieu, 75252 Paris Cedex 05, France}}

\begin{document}
\title{Spectral signature of high-order photon processes enhanced by Cooper-pair pairing}

\author{W.\,C.~Smith}
\thanks{These authors contributed equally to this work. \\ \href{mailto:smithclarke@google.com}{smithclarke@google.com}; Present address: Google Quantum AI, Santa Barbara, CA}
\LPENS
\author{A.~Borgognoni}
\thanks{These authors contributed equally to this work. \\ \href{mailto:smithclarke@google.com}{smithclarke@google.com}; Present address: Google Quantum AI, Santa Barbara, CA}
\LPENS
\author{M.~Villiers}
\LPENS
\author{E.~Roverc'h}
\LPENS
\author{J.~Palomo}
\LPENS
\author{M.\,R.~Delbecq}
\LPENS
\author{T.~Kontos}
\LPENS
\author{P.~Campagne-Ibarcq}
\LPENS
\author{B.~Dou\c{c}ot}
\LPTHE
\author{Z.~Leghtas}
\email[]{zaki.leghtas@ens.fr}
\LPENS
\date{\today}


%
\begin{abstract}
Inducing interactions between individual photons is key for photonic quantum information and studying many-body photon states. Superconducting circuits are well suited to combine strong interactions with low losses. Typically, microwave photons are stored in an $LC$ oscillator shunted by a Josephson junction, where the zero-point phase fluctuations across the junction determine the strength and order of photon interactions. Most superconducting nonlinear oscillators operate with small phase fluctuations, where two-photon Kerr interactions dominate. In our experiment, we shunt a high-impedance $LC$ oscillator with a dipole element favoring the tunneling of paired Cooper pairs. This leads to large phase fluctuations of 3.4, accessing a regime where transition frequencies shift non-monotonically with excitation number. From spectroscopy, we extract two-, three-, and four-photon interaction energies, all of similar strength and exceeding the photon loss rate. Our results open a new regime of high-order photon interactions
in microwave quantum optics.
 
\end{abstract}

\maketitle

\section{Introduction}
{Photons do not interact with each other in free space. In the quantum optical domain, they are typically brought into interaction by coupling them to atoms \cite{Chang2014}. Recent advances have realized two- and three-photon interactions mediated by a dense gas of Rydberg atoms, demonstrating photon dimers and trimers \cite{Liang2018}, and photonic vortices \cite{Drori2023}. Reaching processes of higher order would find applications in multi-photon quantum logic \cite{Prabin2013} and the study of many-body photon states \cite{Greentree2006, Ma2019, Carusotto2020}, but has remained out of reach since it requires inducing even stronger interactions between photons.}


\begin{figure}[t]
\centering
\subfloat{%
\includegraphics{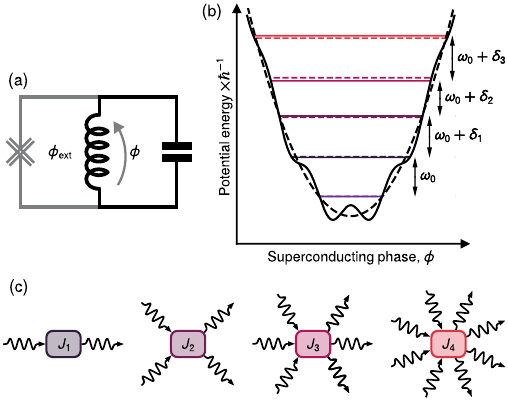}
\label{fig:reduced_circuit}}
\subfloat{\label{fig:potential}}
\subfloat{\label{fig:sketch}}
\caption{Principle of high-order photon processes. (a) Electrical circuit depicting a superconducting $LC$ oscillator (black) shunted by a generalized Josephson junction (grey) that only permits Cooper-pair tunneling in pairs. The circuit is threaded by an external flux denoted $\phi_\text{ext}$. (b) Potential and energy levels (solid lines) of this nonlinear oscillator {[Eq.\ \eqref{eq:H} with parameters $\Omega/2\pi=2.86~\text{GHz},\EJ/h=0.795~\text{GHz}, \eta=3.4$]} and its linear equivalent (dashed lines) as a function of the superconducting phase difference at $\phi_\mathrm{ext} = \pi$. Crucially, the adjacent transition frequency shift $\delta_n$ from the bare frequency $\omega_0$ alternates in sign when ascending the ladder. (c) Photon process diagrams for the first four interaction orders and corresponding interaction energies $J_n$.}
\label{fig:fig1}
\end{figure}


{In the field of microwave quantum optics with superconducting circuits \cite{Wallraff2004, Hofheinz2009, Lang2013}, the nonlinearity of the Josephson junction is employed to mediate interactions between photons. These photons are typically stored in an $LC$ circuit \cite{Devoret1997} of angular frequency $\Omega=1/\sqrt{LC}$ ($L$ and $C$ are the circuit inductance and capacitance, respectively). Superconductivity endows these circuits with low photon loss, and quality factors exceeding one million are routinely observed \cite{Place2021, Ganjam2024, Kono2024}. When such a circuit is shunted by a Josephson junction, interactions between photons appear (Fig.\ \ref{fig:fig1}). The Hamiltonian takes the form
\begin{align}
\bH_\text{ideal} &= \hbar\Omega \ba^\dag \ba-\EJ\cos (\bphi-\phi_\text{ext}) \nonumber\\
\bphi &= \phizpf(\ba+\ba^\dag)\;, 
\label{eq:H}
\end{align}
where $\hbar$ is the reduced Planck constant and $\ba$ is the photon annihilation operator. The interaction energy stems from the Josephson cosine potential with Josephson energy $\EJ$, and $\bp$ is the phase drop across the junction with $\phizpf$ its zero-point fluctuations. The loop formed by the oscillator inductance and the junction is threaded by external magnetic flux denoted $\phi_\text{ext}$. Expanding the cosine into its Taylor series reveals the various interaction processes. For example, the $\left[\phizpf(\ba+\ba^\dag)\right]^4$ term yields a two-photon interaction term $\ba^{\dag^2}\ba^2$ corresponding to the Kerr effect. A celebrated success of microwave quantum optics was the first realization of a Kerr interaction that exceeded the photon loss rate, demonstrating the collapse of a coherent state into multi-component Schr\"odinger cat states \cite{Kirchmair2013}. In this {work}, we address the problem of inducing higher-order processes of the form $\ba^{\dag^n} \ba^n$ where $n=3$, $4$, and beyond.}


\begin{figure}
\centering
\subfloat{%
\includegraphics{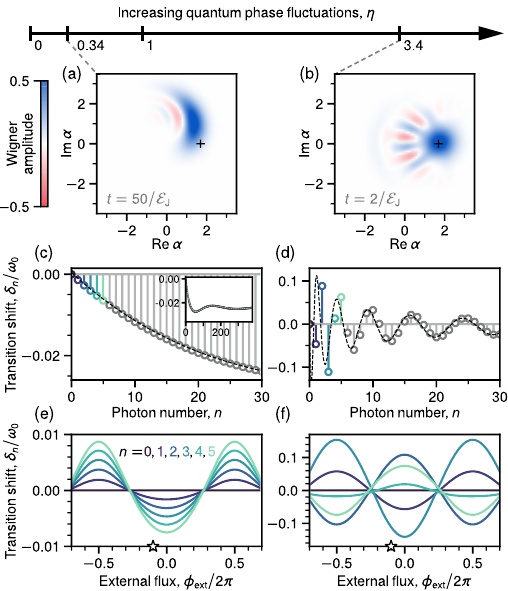}
\label{fig:kerr_wigner}}
\subfloat{\label{fig:displacement_wigner}}
\subfloat{\label{fig:kerr_photons}}
\subfloat{\label{fig:displacement_photons}}
\subfloat{\label{fig:kerr_flux}}
\subfloat{\label{fig:displacement_flux}}
\caption{Signatures of higher-order photon processes. (a)--(b) Simulated Wigner quasiprobability distribution representing the initial time evolution of the coherent state $|\alpha\rangle$ with $\alpha = 1.7$ for small and large quantum phase fluctuations. This value of $\alpha$ was chosen to match $\eta/2$ for $\eta=3.4$ (see text). As quantum phase fluctuations increase, the evolution remarkably transitions from diffusive to nondiffusive. (c)--(f) Simulated transition frequency shifts $\delta_n=\omega_n-\omega_0$, where $\omega_n$ is the transition frequency between energy levels $n$ and $n+1$, represented versus photon number at the starred external flux value (c)--(d), and versus external flux (e)--(f). We observe the transition from an ordered to an alternating arrangement that asymptotically approaches a Bessel function (dashed lines). Simulations correspond to numerical diagonalization and time propagation of Eq.\ \eqref{eq:H} with parameters {$\Omega/2\pi=2.86~\text{GHz},\EJ/h=0.795~\text{GHz}$} and $\eta=0.34$ for (a), (c), (e) and $\eta=3.4$ for (b), (d), (f) as indicated by the top axis.}
\label{fig:fig2}
\end{figure}

{The relative strength of multi-photon processes is governed by the dimensionless quantity $\phizpf$. It can be expressed as $\phizpf=\sqrt{\pi Z/R_Q}$, where $Z=\sqrt{L/C}$ is the $LC$ circuit impedance, and $R_Q\approx \SI{6.4}{\kilo\ohm}$ is the superconducting resistance quantum \cite{Girvin2014}. {The $n$-photon process $\ba^{\dag^n} \ba^n$ has a strength $J_n$ that scales as $\left(\phizpf^n/n!\right)^{2}$. Going beyond the Kerr effect requires that $\lvert J_3/J_2\rvert=\eta/3\gtrsim1$, or equivalently $Z {\gg} R_Q$.} However, fabricating an $LC$ oscillator with a characteristic impedance {far} exceeding the superconducting resistance quantum is challenging. {A successful strategy has been to fabricate the resonator inductance from an array of 40 to 100 Josephson junctions \cite{Manucharyan2009, Pop2014} or a high kinetic inductance material such as granular {aluminum} \cite{Grunhaupt2019}. Values of $\eta\approx 1.8$ have been achieved, giving rise to the {fluxonium} qubit. More recently, {arrays} of 460 Josephson junctions have been suspended above the substrate, achieving $\eta\approx 3.8$, and giving rise to the quasicharge qubit \cite{Pechenezhskiy2020}. Another strategy has been to fabricate the resonator out of a planar coil of thin superconducting wire. Fluctuations of $\phizpf\approx1$ were achieved, and emissions of $k$-photon bunches ($k=1 \text{ to } 6$) were observed by activating the process $\ba^k+\ba^{\dag^k}$ with a voltage-biased junction \cite{Menard2022}. Values as large as $\eta=2.4$ were reported by suspending such a planar coil on a thin substrate \cite{Peruzzo2021}, leading to the observation of phase delocalization with an uninterrupted wire.}

Another route towards large phase fluctuations is to replace the Josephson junction, that allows Cooper-pair tunneling, by a dipole that only allows pairs of Cooper pairs to tunnel \cite{Doucot2002, Ioffe2002.1}. In the basis of tunneled Cooper-pair number $N$, the tunneling operator is transformed as  $\frac{1}{2}\sum_N{\left(\ket{N}\bra{N+1}+\ket{N+1}\bra{N}\right)} \rightarrow \frac{1}{2}\sum_N{\left(\ket{N}\bra{N+2}+\ket{N+2}\bra{N}\right)}$. Equivalently, in the {conjugate} phase representation $\varphi$, $\cos\bvarp\rightarrow \cos2\bvarp$. {Shunting such an element by an $LC$ oscillator [Fig.\ \subref*{fig:reduced_circuit}],}{ and denoting $\bp=2\times\bvarp$, we see that phase fluctuations are effectively doubled: $\phizpf=2\times \sqrt{\pi Z/R_Q}$ \cite{Smith2020}}. In the extreme regime of {$\phizpf \gtrsim 3$}, we further require that $\EJ\ll\hbar\Omega$ [Fig.\ \subref*{fig:potential}], so that the Josephson cosine potential primarily induces $n$-photon interaction processes $\ba^{\dag^n} \ba^n$ [Fig.\ \subref*{fig:sketch}].}


\begin{figure}
\centering
\subfloat{%
\includegraphics{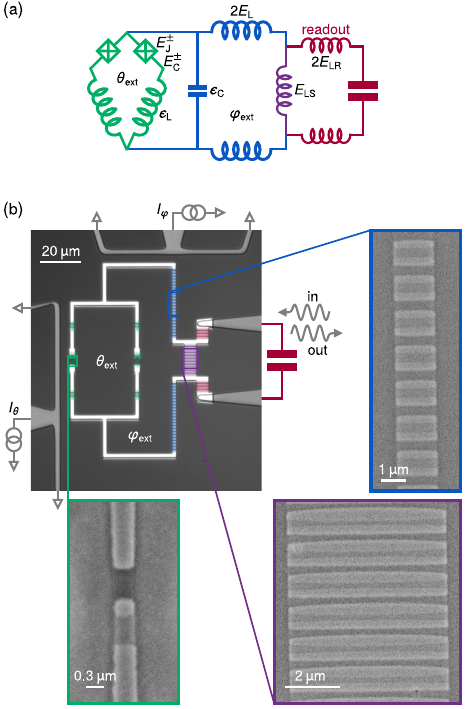}
\label{fig:circuit_kite}}
\subfloat{\label{fig:device}}
\caption{Experimental implementation. (a) Lumped element circuit of the $LC$ oscillator (blue) shunted by the KITE (green) and coupled through a shared inductance (purple) to a readout mode (red). The two small junctions have slightly different Josephson energies $E_J^\pm = (1 \pm \varepsilon) E_J$ and charging energies $E_C^\pm = E_C / (1 \pm \varepsilon)$, with $\varepsilon \ll 1$, due to junction fabrication variation. (b) Optical micrograph of the physical device, with false color indicating the constituent Josephson junctions {and their respective scanning electron microscope images (from a nominally identical sample).} Aluminum and niobium electrodes appear in white and grey, respectively. {(Green {frame}) one of the two small KITE junctions. (Blue {frame}) $14$ of the array junctions that form both the internal KITE inductance and the inductive shunt. (Purple {frame}) $12$ of the array junctions that form the shared inductance between the circuit and the readout resonator, as well as the self-inductance of the readout. All junctions are fabricated in one step using Dolan bridges.}}
\label{fig:fig3}
\end{figure}


\begin{figure*}[t]
\centering
\subfloat{%
\includegraphics{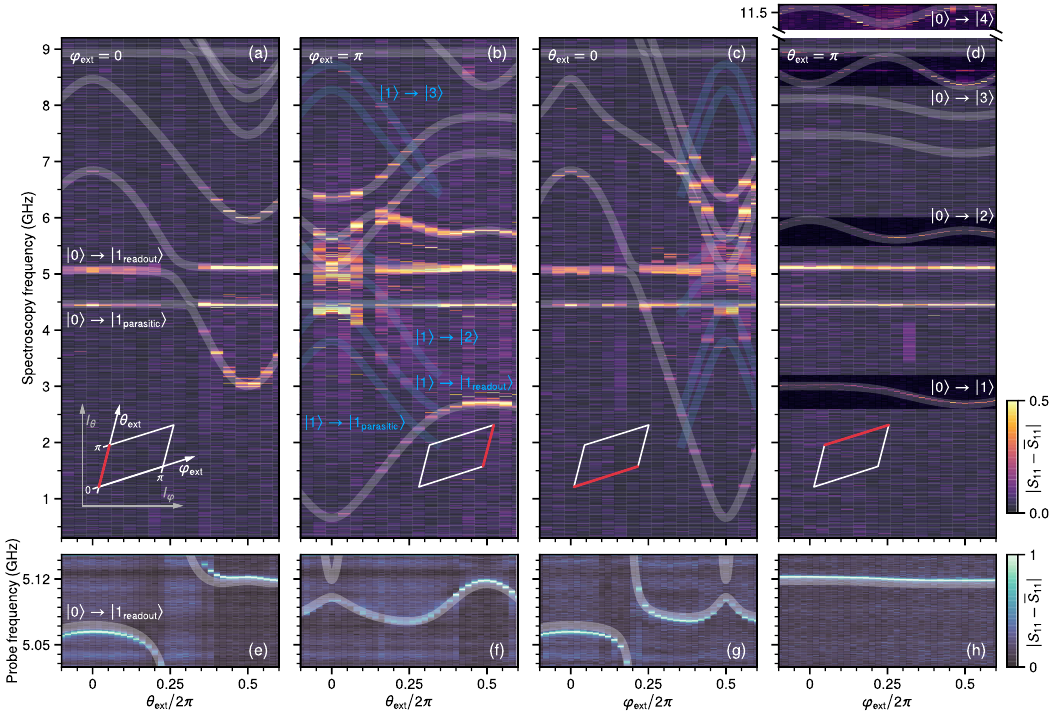}
\label{fig:twotone_theta_zero}}
\subfloat{\label{fig:twotone_theta_half}}
\subfloat{\label{fig:twotone_varphi_zero}}
\subfloat{\label{fig:twotone_varphi_half}}
\subfloat{\label{fig:readout_theta_zero}}
\subfloat{\label{fig:readout_theta_half}}
\subfloat{\label{fig:readout_varphi_zero}}
\subfloat{\label{fig:readout_varphi_half}}
\caption{{Spectroscopy measurements.} (a)--(d) Two-tone reflection spectroscopy (background subtracted) of {the lowest transitions of the circuit} along the four edges of the primitive cell in the two-dimensional external flux landscape (inset diagrams), and (e)--(h) accompanying readout spectroscopy. Theoretical transition frequencies {from the ground state (semi-transparent white lines)} are obtained from numerical diagonalization of the {five}-mode circuit model with {seven} fitted Hamiltonian parameters {(see Tab.\ \protect\ref{tab:params})}. {Additionally, transition frequencies from the first excited state (semi-transparent blue lines) are shown when the first excited state has a transition frequency that falls below $\SI{2.5}{\giga\hertz}$ and is therefore thermally occupied.} Note that this model {includes} the nearly harmonic readout {and} parasitic modes [labeled in (a)]. {Scans around the $\ket{0}\rightarrow \ket{n}$ transitions for $n=1-4$ [labeled in (d)] are enlarged in Fig.~\ref{fig:spectroscopy_zoom_cuts}.}}
\label{fig:fig4}
\end{figure*}
{In this experiment, we implement a superconducting $LC$ oscillator shunted by {an approximate two-Cooper-pair tunneling} element. {We place ourselves in the unexplored regime where the tunneling energy is smaller than the oscillator transition energies, and photon-photon interactions of order larger than two (Kerr) dominate, or equivalently zero-point phase fluctuations exceed 3. We achieve $\EJ/\hbar\Omega = 0.28$ and $\phizpf = 3.4$}. {We probe our circuit through microwave spectroscopy, which is better suited than correlation measurements to the regime where interactions exceed the oscillator linewidth}. We measure the first four transition energies of our device, and find that unlike a Kerr resonator, they do not follow a monotonic trend. Instead,  we observe an alternation of the sign of the oscillator frequency shift for each added photon. From this spectroscopic signature, we extract {two-, three-, and four-photon} interaction processes of amplitudes greater than {\SI{70}{\mega\hertz}}, that alternate in sign, and far exceed the transition linewidths of {\SI{200}{\kilo\hertz}}. {Entering the regime of strong high-order photon interactions opens many possibilities in microwave quantum optics such as multi-photon quantum logic \cite{Prabin2013}, the study of many-body photon states \cite{Ma2019}, or the processing of protected qubits \cite{Cohen2017}.}}

\section{Toy model}
\label{sec:toymodel}
We proceed to the analysis of the ideal Hamiltonian in Eq.\ \eqref{eq:H} in the regime $\EJ\ll\hbar\Omega$ and $\phizpf>1   $ (Fig.\ \ref{fig:fig1}), and for simplicity, we set $\phi_\text{ext}=0$. Note that the cosine in Eq.\ \eqref{eq:H} may be decomposed as $\cos\left[\phizpf(\ba+\ba^\dag)\right]=\frac{1}{2}\left(\bD_{\phizpf}+\bD_{-\phizpf}\right)$, where $\bD_{\phizpf}=\exp\left[i\phizpf(\ba+\ba^\dag)\right]$ is the displacement evolution operator. Remarkably, this evolution operator that usually results from the integration of a linear Hamiltonian $\propto \ba+\ba^\dag$ over time enters the Hamiltonian directly. As a consequence, even at short times, a quantum state evolving under the Hamiltonian in Eq.\ \eqref{eq:H} will be displaced across phase-space by $\pm\phizpf$. The effect is particularly striking when initializing the system in a coherent state of amplitude $\alpha=\eta/2$  \cite{Cohen2017}, so that $\ket{\pm\alpha}$, that are distant by $2\alpha$ in phase-space, are directly coupled through $\cos\left[\phizpf(\ba+\ba^\dag)\right]$ [Fig.\ \subref*{fig:displacement_wigner}]. This is qualitatively different from the familiar diffusive-like evolutions resulting from low-order photon interactions \cite{Kirchmair2013} [Fig.\ \subref*{fig:kerr_wigner}]. 

We now express $\hat{\mathcal{H}}_\text{ideal}$ in Eq.\ \eqref{eq:H} in terms of $n$-photon interaction processes [Fig.\ \subref*{fig:sketch}]. We start by expanding the cosine into a normal ordered Taylor expansion. {Since $\EJ\ll\hbar\Omega$}, by virtue of the rotating wave approximation (RWA), we neglect non-particle number conserving terms. We arrive at \cite{supmat} :
\begin{equation}
\label{eq:HJn}
\bH_\text{ideal}\approx\hbar\omega_0 \ba^\dag \ba+\sum_{n\ge2}J_n \ba^{\dag^n}\ba^n\;,
\end{equation}
where the $n$-photon interaction energy takes the form $J_n=-\EJ e^{-{\phizpf^2}/2}(-1)^n\left({\phizpf^n}/n!\right)^2$, and the renormalized frequency is $\omega_0=\Omega+ J_1/\hbar$. Note that $J_{n}/J_{n-1}=-(\phizpf/n)^2$, and hence the interaction strength is maximal for the {integer order closest to $\phizpf$.} 
{The eigenstates of Hamiltonian~\eqref{eq:HJn} are Fock states} with eigenenergies $E_n=n\hbar \omega_0+\sum_{k=2}^n\frac{n!}{(n-k)!}J_k$, and the experimentally accessible quantities are the transition frequencies $\omega_n=(E_{n+1}-E_n)/\hbar$. We introduce the transition frequency shift in the presence of $n$ photons as $\delta_n=\omega_n-\omega_0$ 
[Fig.\ \subref*{fig:potential}] and we find :
\begin{equation}
\label{eq:deltanJn}
\delta_n=\sum_{k=2}^{n+1} k \frac{n!}{(n+1-k)!} J_k/\hbar\;.    
\end{equation}
In the familiar situation of the Kerr oscillator where $\phizpf\ll1$, $J_2\approx -\EJ\phizpf^4/4$ is half the Kerr shift per photon and $J_{n\ge 3}$ can be neglected. Hence $\omega_n \approx \omega_0 + 2 n J_2/\hbar$ and the transition frequency monotonically shifts for each added photon [Figs.\ \subref*{fig:kerr_photons} and \subref*{fig:kerr_flux}]. This is in stark contrast with the regime of extreme phase fluctuations explored in this work where the transition frequency shift may alternate in sign for each added photon [Figs.\ \subref*{fig:displacement_photons} and \subref*{fig:displacement_flux}]. This resembles the oscillatory nonlinearity predicted in a resonator containing a phase-slip element \cite{Hriscu2011}.

\section{Circuit implementation
}
Our circuit implementation of $\hat{\mathcal{H}}_\text{ideal}$ in Eq.\ \eqref{eq:H} is depicted in Figs.\ \subref*{fig:circuit_kite} and \subref*{fig:device}. It consists of a high impedance $LC$ oscillator. The inductance, which we aim to maximize, is formed by a chain of {$109$} Josephson junctions, {19 of which are shared with a readout resonator (inductive energy $E_{LS}/h=\SI{11.91}{\giga\hertz}$) and 90 unshared junctions (inductive energy $E_L/h=\SI{0.57}{\giga\hertz}$)}. Note that approximating the junction chain inductance by a single inductor is only valid at frequencies lower than the first chain mode (that we estimate above 10 GHz). {The capacitance, which we aim to minimize, has multiple contributions. {The first one is the self capacitances of the small junctions in the} tunneling element (described below) attached to the chain of junctions. The second one arises from the capacitance between the two wires linking the chain of junctions to the tunneling element, resulting in a charging energy $\epsilon_C/h={\SI{3.24}{\giga\hertz}}$ (estimated from {finite-element} simulations). {Other sources} of capacitive loading, not accounted for in our model, {are from the self capacitance and capacitance to ground of the chain junctions} \cite{Viola2015}.}

{Tunneling occurs through a so-called Kinetic Interference coTunneling Element (KITE) \cite{Bell2014, Smith2020}}. It consists of two parallel arms that form a loop threaded by {an} external flux $\thetaext$. {Each arm contains a small junction of Josephson energy $E_J^{\pm}=E_J(1\pm\epsilon)$ and charging energy $E_C^\pm=E_C/(1\pm\epsilon)$, with $E_J/h=\SI{3.98}{\giga\hertz}$, $E_C/h=\SI{10.40}{\giga\hertz}$ and fabrication uncertainty results in a small asymmetry factor $\epsilon=0.033$. Each small junction is placed in series with $6$ large junctions of a total inductive energy $\epsilon_L/h=\SI{6.16}{\giga\hertz}$. 
Additionally, our chip contains a lumped $LC$ readout resonator composed of two planar capacitor pads {[}not shown in Fig.\ \subref*{fig:device}{]} and an array of 29 junctions, 19 of which are shared with the main circuit, and 10 unshared junctions (inductive energy $E_{LR}/h=\SI{18.03}{\giga\hertz}$). Through the small KITE junctions, this inductive coupling induces a dispersive interaction between the circuit and its readout resonator \cite{Smith2016}. Finally, our circuit hosts a ``parasitic'' mode, visible in electromagnetic simulations, where current flows symmetrically through both halves of the junction chain (with minimal current in the shared junctions), charging and discharging a capacitor {[}not represented in Fig.\ \subref*{fig:circuit_kite}{]} formed between the readout pads and the connecting leads to the KITE.}



The circuit parameters quoted above are extracted by fitting a {five}-mode circuit model \cite{supmat} {(including the readout and parasitic modes)} to two-tone spectroscopy data at various flux biases $(\thetaext, \phiext)$ [Figs.\ \subref*{fig:twotone_theta_zero}--\subref{fig:twotone_varphi_half}]. This {five}-mode Hamiltonian is $2\pi$ periodic in $(\thetaext, \phiext)$ and its spectrum possesses inversion symmetry about $(\thetaext, \phiext)=(0,0)$, $(0,\pi)$, $(\pi,0)$, and $(\pi,\pi)$ \cite{Smith2022}. Therefore, these four points are the vertices of a plaquette that constitutes the primitive cell of the circuit spectrum as a function of external flux. We acquire the circuit spectrum along the edges of this plaquette [diagrams in Figs.\ \subref*{fig:twotone_theta_zero}--\subref{fig:twotone_varphi_half}]. At each bias point, we start by acquiring the reflection spectrum of the readout resonator [Figs.\ \subref*{fig:readout_theta_zero}--\subref{fig:readout_varphi_half}]. We then set the readout tone on resonance, and sweep a probe tone over a broad spectral range [Figs.\ \subref*{fig:twotone_theta_zero}--\subref{fig:twotone_varphi_half}]. When the probe hits a circuit transition from the {low lying states} to the $n$-th excited state, the reflected readout signal is affected. We identify several transitions that are captured by a {five}-mode circuit model \cite{supmat} ({semi-transparent lines}). {The features near $\SI{5.1}{\giga\hertz}$ and $\SI{4.4}{\giga\hertz}$ correspond to the readout and parasitic modes, respectively}. The circuit parameters we extract from this fit are summarized in Tab.\ \ref{tab:params}.


\begin{table}
\centering
\begin{tabular}{c c c c c c c | c}
\hline
\hline
\rule{0pt}{2.5ex} $E_J/h$ & $E_C/h$ & $E_L/h$ & $\epsilon_L/h$ & $\varepsilon$ & $E_{LR}/h$ & $E_{LS}/h$ & {$\epsilon_C/h$} \\
\hline
\rule{0pt}{2.5ex} 3.98 & 10.40 & 0.57 & 6.16 & 0.033 & 18.03 & 11.91 & {3.24} \\
\hline
\hline
\end{tabular}
\caption{{Device parameters corresponding to the five-mode circuit in Fig.\ \protect\subref*{fig:circuit_kite}. The first seven parameters are found by fitting the spectral lines in Figs.\ \protect\subref*{fig:twotone_theta_zero}--\protect\subref{fig:readout_varphi_half} to the circuit Hamiltonian, with all capacitances (except junction capacitances) fixed to values from finite-element simulations \cite{supmat}. The final parameter---the oscillator charging energy in the absence of the junctions---is computed from the full device capacitance matrix and is not found from the fit.}\label{tab:params}}
\end{table}

\begin{table}
\centering
\begin{tabular}{c|c c c c c}
\hline
\hline
\rule{0pt}{2.5ex} $\thetaext$ & $\Omega/2\pi$ & $E_{J1}/h$ & $E_{J2}/h$ & $\varphi_\text{zpf}$ & $2\varphi_\text{zpf}$ \\
\hline
\rule{0pt}{2.5ex} 0 & {2.95} & {5.31} & {0.658} & {1.67} & {3.34} \\
\hline
\rule{0pt}{2.5ex} $\pi$ & {2.86} & {0.27} & {0.795} & {1.70} & {3.40}\\
\hline
\hline
\end{tabular}
\caption{{Extracted one-mode model parameters found by fitting the measured transition frequencies at $\theta_\text{ext}=0, \pi$ (Fig.\ \protect\ref{fig:fig5}) to the effective one-mode Hamiltonian in Eq.\ \eqref{eq:Hcircuit}. All energy scales are given in gigahertz.} \label{tab:params2}}
\end{table}


\section{High-order photon processes}
We now delve into explaining how our circuit emulates the Hamiltonian of Eq.\ \eqref{eq:H}. {Here, we will provide an intuitive understanding of this circuit, and refer the reader to the appendices for a more rigorous analysis.} We start by discarding the readout mode, the parasitic mode, and the two KITE self-resonant high-frequency modes, and focus on the $LC$ oscillator. {Since $E_{LS}\gg E_L$, the total oscillator inductive energy is {approximately} $E_L$. The total oscillator charging energy is $\epsilon_{C,{\mathrm{tot}}}=1/(2/E_C+1/\epsilon_{{C}})=h\times {\SI{2.0}{\giga\hertz}}$}. {The resonant frequency of this oscillator is $\Omega\approx\sqrt{8E_L\epsilon_{C,{\mathrm{tot}}}}/\hbar$.} In the regime $\epsilon_L\gg E_J$, the potential energy of one arm of the KITE traversed by a phase drop of $\varphi$ takes the form $U^\pm(\varphi)\approx-E^\pm_J\cos\varphi+\frac{{E_J^\pm}^2}{4\epsilon_L}\cos2\varphi$ and higher harmonics have been neglected \cite{supmat}. {Biasing the circuit at $\thetaext=0$, both Cooper-pair tunneling and cotunneling across both arms interfere constructively. Indeed, the potential energy of the KITE is $U^+(\varphi)+U^-(\varphi)\approx- 2E_J\cos\varphi+\frac{{E_J}^2}{2\epsilon_L}\cos2\varphi$.} Biasing the circuit at $\thetaext=\pi$, Cooper-pair tunneling across both arms interferes destructively, while cotunneling interferes constructively. Indeed, the potential energy of the KITE is $U^+(\varphi)+U^-(\varphi+\pi)\approx-2\epsilon E_J\cos\varphi+\frac{E_J^2}{2\epsilon_L}\cos2\varphi$. In summary, this yields an effective Hamiltonian for our circuit of the form \cite{supmat}:
\begin{align}
\label{eq:Hcircuit}
\hat{\mathcal{H}}_\text{circuit}({\thetaext})&=\hbar\Omega\ba^\dag\ba-E_{J1}({\thetaext})\cos(\bvarp-\phiext)\nonumber\\
&\qquad\qquad\;+E_{J2}\cos[2(\bvarp-\phiext)],
\end{align}
where $\phiext$ is the flux threading the loop formed by the KITE and the oscillator inductance {and the phase operator verifies $\hat\varphi=\varphizpf\left(\ba + \ba^\dag \right)$, where {$\varphizpf=(2\epsilon_{C,{\mathrm{tot}}}/E_L)^{1/4}$} is the zero-point phase fluctuations}.

\begin{figure}[t]
\centering
\subfloat{%
\includegraphics{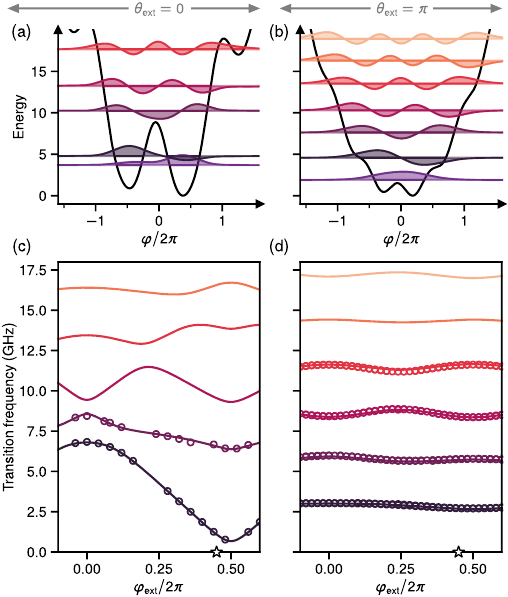}
\label{fig:potential_zero}}
\subfloat{\label{fig:potential_half}}
\subfloat{\label{fig:transitions_zero}}
\subfloat{\label{fig:transitions_half}}
\caption{{Switching our circuit between a fluxonium and an oscillator with high order interactions. (a)--(b) Potential energy $U(\varphi)=\frac{1}{2} E_L \varphi^2 - E_{J1}\cos(\varphi-\phiext) + E_{J2}\cos[2(\varphi-\phiext)]$ as a function of superconducting phase at $\phiext=0.9\pi$ [starred in (c)--(d)], for (a) $\thetaext=0$ and (b) $\thetaext=\pi$. (c)--(d) Transition frequencies from the ground state as a function of external flux. The data points (open circles) correspond to the {average of resonances visible in scans such as} Figs.\ \protect\subref*{fig:twotone_varphi_zero}--\protect\subref{fig:twotone_varphi_half}. Theoretical transition energies (solid lines) are obtained from Hamiltonian \eqref{eq:Hcircuit} with fitted parameters reported in Tab.\ \ref{tab:params2}.}}
\label{fig:fig5}
\end{figure}

{Let us now analyze the case where $\thetaext=0$ [Figs.\ \subref*{fig:potential_zero} and \subref*{fig:transitions_zero}]. From the simple theory sketched above, we expect $E_{J1}(\thetaext=0)\approx2E_J$ and $E_{J2}(\thetaext=0)\approx E_J^2/2\epsilon_L$. In the regime of this experiment where $E_J\ll \epsilon_L$, we have $E_{J2}\ll E_{J1}$, and so the dominant term in the potential is the regular Cooper-pair tunneling energy. Recall that the term in $\Omega \ba^\dagger\ba$ may be equivalently recast as {$4\epsilon_{C,{\mathrm{tot}}} \hat{N}^2+\frac{1}{2}E_L \hat\varphi^2$}, where $\hat N$ is the Cooper-pair number operator conjugate to $\hat\varphi$. Interestingly, our experiment is in the regime where {$E_{J1}\gtrsim \epsilon_{C,{\mathrm{tot}}}\gg E_L$}, which is typical of a fluxonium \cite{Manucharyan2009}. The fluxonium is nothing like an anharmonic oscillator. Indeed, the fluxonium eigenstates include fluxon states pinned in Josephson wells [Fig.\ \subref*{fig:potential_zero}], and that therefore strongly disperse with flux [Fig.\ \subref*{fig:transitions_zero}], and plasmon states that are weakly flux-dependent. An anharmonic oscillator only has weakly flux-sensitive plasmon states. Consequently, the language of interacting photons is not adapted to describe a fluxonium.}

{We now turn to the case where $\thetaext=\pi$ [Figs.\ \subref*{fig:potential_half} and \subref*{fig:transitions_half}]. From the simple theory sketched above, we expect $E_{J1}(\thetaext=\pi)\approx2\epsilon E_J$ and $E_{J2}(\thetaext=\pi)\approx E_J^2/2\epsilon_L$. Conveniently, for sufficiently symmetrical junctions and adequately choosing $E_J\ll\epsilon_L$, we may enter the regime where $E_{J1}\ll E_{J2}\ll\Omega$. In this regime, the potential is dominated by Cooper-pair cotunneling, while regular tunneling may be considered an undesired perturbation. In addition, these nonlinear terms are smaller than the linear oscillator term $\Omega\ba^\dagger\ba$. Consequently, this system is well described by a nonlinear oscillator of quasi-equally spaced photonic Fock states that couple through Cooper-pair cotunneling [Figs.\ \subref*{fig:potential_half} and \subref*{fig:transitions_half}]. The extent of the zero-point phase fluctuations is visible through the number of Josephson corrugations covered by the ground state wave-function.} We may now establish a correspondence with the ideal Hamiltonian of Eq.\ \eqref{eq:H}. Note that $\hat{\mathcal{H}}_\text{circuit}(\theta_\text{ext}=\pi)$ corresponds to $\hat{\mathcal{H}}_\text{ideal}$ up to the perturbative term in $E_{J1}$, with the correspondence $\EJ=E_{J2}$, $\bp=2\bvarp$ and hence $\phizpf=2\times\varphizpf$ and $\phi_\text{ext}=2\varphi_\text{ext}+\pi$. Additionally, the $J_n$ obtained from Eq.\ \eqref{eq:Hcircuit} depend on external flux and contain an added contribution from the term in $E_{J1}$.

{The ability to switch our device in-situ from a familiar fluxonium-like circuit to an original nonlinear oscillator endowed with high-order interactions is convenient to benchmark our system. We extract the parameters of the one-mode Hamiltonian in Eq.\ \eqref{eq:Hcircuit} by fitting this model to the measured $\phiext$-dependent transition energies at $\theta_\text{ext}=0$ and $\pi$ (Fig.\ \ref{fig:fig5}). The resulting parameters are displayed in Tab.\ \ref{tab:params2}. For each $\phiext$-dependent dataset, we perform a four-parameter fit ($\Omega, E_{J1}, E_{J2}, \varphizpf$). The fits converge on two values of $\Omega$ that are within $3\%$ of each other,  $E_{J2}$ within $20\%$ and  $\varphizpf$ within $2\%$. This is consistent with the prediction that these parameters should be the same at $\theta_\text{ext}=0$ and $\pi$. On the other hand, the fit converges on two very different values for $E_{J1}$. Indeed, its value at $\theta_\text{ext}=\pi$ is 20 times smaller than the one at $\theta_\text{ext}=0$. This is consistent with our understanding that regular Cooper-pair tunneling constructively interferes at $\theta_\text{ext}=0$, while it destructively interferes at $\theta_\text{ext}=\pi$.}

\begin{figure}[t]
\centering
\subfloat{%
\includegraphics{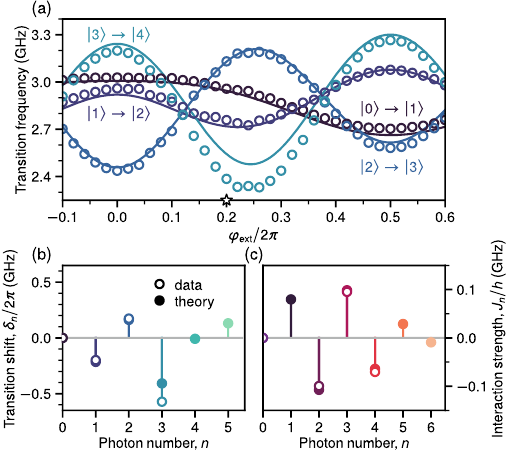}
\label{fig:interlacing}}
\subfloat{\label{fig:Kn}}
\subfloat{\label{fig:Jn}}
\caption{Spectral interlacing. (a) Adjacent transition frequencies obtained from two-tone measurements (open circles) and numerical diagonalization of the one-mode Hamiltonian of Eq.\ \eqref{eq:Hcircuit} (solid curves) along $\theta_\mathrm{ext} = \pi$. (b) Transition frequency shifts $\delta_n$ and (c) interaction strengths $J_n$ extracted from measurements (open circles) and analytic expressions depending on the fitted circuit parameters (full circles) at the starred external flux value. The transition frequency shifts alternate in sign while remaining much smaller than the transition frequency itself, directly corresponding to similarly-sized Hamiltonian coefficients for two-, three-, and four-photon interaction strengths.}
\label{fig:fig6}
\end{figure}

{Finally, we focus on the case $\theta_\text{ext}=\pi$ in order to extract multi-photon interaction strengths from the measured transition frequencies (Fig.\ \ref{fig:fig6})}. Two notable features are visible in the data. First, as previously discussed, these transition frequencies vary in a $\pm \SI{500}{\mega\hertz}$ window---a $\pm \SI{17}{\percent}$ fraction of the central frequency $\Omega/2\pi=\SI{2.86}{\giga\hertz}$. This confirms that the flux-dependent tunneling amplitude is a perturbation to the $LC$ oscillator frequency, {i.e.\ $E_{J1}, E_{J2} <\hbar\Omega$. In particular, we find $E_{J2}/h=\SI{0.795}{\giga\hertz}$ and the perturbation $E_{J1}/h=\SI{0.27}{\giga\hertz}$}. Second, the transition frequencies $\omega_n$ between levels $n$  and $n+1$ are not ordered in $n$. Instead, they interlace as a function of $\phiext$, indicating that we have entered the regime of large phase fluctuations. For example, at {$\phiext/2\pi=0.2$}, $\omega_0>\omega_1$, $\omega_1<\omega_2$, and $\omega_2>\omega_3$ [Fig.\ \subref*{fig:Kn}]. From this measured spectrum, we compute the $n$-photon interaction strengths $J_n$ for $n=2$, $3$, and $4$ [Fig.\ \subref*{fig:Jn}] by inverting Eq.\ \eqref{eq:deltanJn} \footnote{Notice that $J_1$ is experimentally inaccessible since it corresponds to the shift between the measured transition frequency $\omega_0$ and the $LC$ resonance $\Omega$ in the absence of the tunneling element}. Notably, we find that $|J_2|\approx |J_3|$, which is consistent with the extracted {$2\varphizpf = 3.4$}.

\section{Discussion}
In conclusion, this experiment explores a new regime of nonlinear microwave quantum  optics where interactions between photons are so strong that second-, third-, and fourth-order processes are of comparable amplitude and largely exceed the photon decay rate. We access this regime with photons stored in a high impedance $LC$ oscillator that is shunted by a two-Cooper-pair tunneling element, effectively boosting phase fluctuations. Two technical challenges must be met: the tunneling energy {$E_{J2}$} must be weaker than the oscillator energy $\hbar\Omega$, and the {boosted phase fluctuations $2\varphizpf$} across the tunneling element must exceed {3}. We measure the first four transition frequencies of our circuit, and observe their interlacing versus flux. From these spectra, we extract {$E_{J2}/\hbar\Omega = 0.28$ and $2\varphizpf = 3.4$}.

This experiment could be extended in multiple ways. First, one could improve the quantitative analysis by improving the spectroscopic data (larger frequency spans, denser flux sweeps and more averaging). Another direction could be the study of the quantum dynamics and scattered radiation correlations of this system under the action of drives and dissipation  \cite{Hriscu2011}. Moreover, coupling our two-Cooper-pair tunneling element to an array of resonators could induce high-order interactions between multiple modes, useful for the study of many-body photon states \cite{Greentree2006, Ma2019}. Finally, applications are envisioned to process quantum information that is encoded non-locally over the phase space of an oscillator \cite{Cohen2017}.

\vspace{1cm}
\textbf{Ethics declaration:}
The authors declare no competing interests.

\textbf{Data availability:}
The data that support the findings of this work are available from the corresponding author upon reasonable request.

\textbf{Code availability:}
The code used for data acquisition, analysis and visualization is available from the corresponding author upon reasonable request.

\textbf{Author contribution:} W.C.S.\ and Z.L.\ conceived the experiment. W.C.S.\ and A.B.\ designed and measured the device. A.B.\ fabricated the sample. W.C.S, A.B. and Z.L.\ analyzed the data. M.V, J.P, M.R.D, T.K.\ and P.C-I.\ provided experimental support. E.R.\ and B.D.\ provided theory support. W.C.S, A.B.\ and Z.L.\ co-wrote the manuscript with input from all authors.

\textbf{Acknowledgments:} W.C.S acknowledges fruitful discussions with Agustin Di Paolo. The devices were fabricated within the consortium Salle Blanche Paris Centre. We thank Jean-Loup Smirr and the Collège de France for providing nano-fabrication facilities. This work was supported by the QuantERA grant QuCOS, by ANR 19-QUAN-0006-04. This project has received funding from the European Research Council (ERC) under the European Union’s Horizon 2020 research and innovation program (grant agreement no.\ 851740). This work has been funded by the French grants ANR-22-PETQ-0003 and ANR-22-PETQ-0006 under the ``France 2030 Plan.''

\bibliography{biblio}

\renewcommand{\thepage}{S\arabic{page}} 
\renewcommand{\thesection}{S\arabic{section}}  
\renewcommand{\thetable}{S\arabic{table}}  
\renewcommand{\thefigure}{S\arabic{figure}}

\def\theequation{S\arabic{equation}}

\begin{center}
\textbf{\Large Supplementary information}
\end{center}

\section{Device fabrication}
This section details the fabrication process we follow to produce the sample of this experiment.

\paragraph{Wafer preparation:}
The circuit is fabricated on a $\SI{430}{\micro\meter}$-thick wafer of 0001-oriented, double-side epi-polished Sapphire C. The sapphire wafer is initially cleaned through a stripping process in a reactive ion etching (RIE) machine, after which it is loaded into a sputtering system. After one night of pumping, we initiate an argon milling cleaning step, followed by the sputtering of $\SI{120}{\nano\meter}$ of niobium. Subsequently, we apply a protective layer of poly(methyl methacrylate) (PMMA A6), dice the wafer and clean the small chips in solvents. This is followed by a $\SI{2}{\minute}$ oxygen-stripping process and approximately $\SI{30}{\second}$ of exposure to sulphur hexafluoride (SF6) in order to remove the oxide layer formed on the niobium during the stripping process.

\paragraph{Circuit patterning:} We spin optical resist (S1805) and pattern the large features (control lines and readout resonator capacitor pads) using a laser writer. After development (MF319), we rinse in de-ionized water for $\SI{1}{\minute}$, and etch the sample in SF6 with a $\SI{20}{\second}$ over-etch. Finally, the sample is cleaned for $\SI{10}{\minute}$ in acetone at $\SI{50}{\celsius}$. 

\paragraph{Junction patterning: } Next, we apply a bilayer of methacrylic acid/methyl methacrylate [MMA (8.5) MAA EL10] and poly(methyl
methacrylate) (PMMA A6). The entire circuit (KITE, inductive shunt, and readout resonator) is patterned in a single e-beam lithography step. The development takes place in a 3:1 isopropyl alcohol (IPA)/water solution at $\SI{6}{\celsius}$ for $\SI{90}{\second}$, followed by $\SI{10}{\second}$ in IPA. The undercut regions of the bilayer are cleaned by oxygen-stripping for $\SI{30}{\second}$.

\paragraph{Junction deposition: } The chip is then loaded in an e-beam evaporator. We start with a thorough argon ion milling for $\SI{2}{\minute}$ at $\SI{\pm 30}{\degree}$ angles. We then evaporate $\SI{35}{\nano\meter}$ and $\SI{100}{\nano\meter}$ of aluminum, at $\SI{\pm30}{\degree}$ angles, separated by an oxidation step in $\SI{200}{\milli\bar}$ of pure oxygen for $\SI{10}{\minute}$.

\paragraph{Junction characteristics: }
The Josephson junctions are all fabricated from Al/AlOx/Al in a single evaporation step, utilizing the Dolan bridge method. The e-beam base dose is set to $\SI{283}{\micro\coulomb\per\square\centi\meter}$, with an acceleration voltage of $\SI{20}{\kilo\volt}$ and a lens aperture of $\SI{7.5}{\micro\meter}$. Three types of junctions are fabricated. (i) Two small junctions are located in the KITE, with an area of $\SI{0.076}{\square\micro\meter}$. They are patterned with a dose factor of $0.9$ and an undercut dose of $0.2$, resulting in an inductance per junction of $\SI{42}{\nano\henry}$. (ii) A total of $12$ large array junctions are located in the KITE loop, along with another $90$ unshared shunting array junctions. These junctions have an area of $\SI{0.62}{\square\micro\meter}$, and are patterned with a dose factor of $0.9$ and an undercut dose of $0.1$, resulting in an inductance per junction of $\SI{4}{\nano\henry}$. (iii) There are $29$ larger array junctions that constitute the readout resonator inductance, $19$ of which are shared. These junctions have an area of $\SI{3}{\square\micro\meter}$, and are patterned with a dose factor of $0.8$ and an undercut dose of $0.1$, resulting in an inductance per junction of $\SI{0.7}{\nano\henry}$.

\paragraph{Sample mounting: } The chip was subsequently glued with PMMA onto a PCB, wire-bonded and mounted into a sample holder. The device was then thermally anchored to the base plate of a Bluefors dilution refrigerator, surrounded by three concentric cans for magnetic and infrared shielding (outer: cryoperm, middle: aluminum, inner: copper). An optical micrograph of the circuit and SEM images of some junctions are shown in {Fig.\ \ref{fig:device}}.



\section{Mathematical derivations \label{sec:math}}

In this section we derive the expression of the $n$-photon interaction strength $J_n$ that enters Eq.\ \eqref{eq:HJn}, as well as Eq.\ \eqref{eq:deltanJn} that relates $J_n$ to the transition frequency shift $\delta_n$.

We start from Hamiltonian~\eqref{eq:H}  with $\phi_\text{ext}=0$ for simplicity. Note that the cosine may be decomposed as $\cos\left[\phizpf(\ba+\ba^\dag)\right]=\frac{1}{2}\left(\exp\left[i\phizpf(\ba+\ba^\dag)\right]+\exp\left[-i\phizpf(\ba+\ba^\dag)\right]\right)$. Since the commutator $[\ba, \ba^\dag] = 1$ commutes with both $\ba$ and $\ba^\dag$, we may use Glauber's formula as follows $\exp\left[i\phizpf(\ba+\ba^\dag)\right]=e^{-\eta^2/2} e^{i\eta\ba^\dag}e^{i\eta\ba}$. We then expand each exponential into its Taylor series. Since we place ourselves in the regime $\hbar\Omega\gg \mathcal{E}_J$, we perform the rotating wave approximation (RWA) by neglecting terms that do not conserve particle number, yielding 
\begin{equation*}
\cos\bp\biggr \rvert_\text{RWA}=
e^{-\phizpf^2/2}\sum_{n=0}^{+\infty}(-1)^n\left(\frac{\phizpf^{n}}{n!}\right)^2\ba^{\dag^n}\ba^n\;.
\end{equation*}
Inserting this last equation in Eq.\ \eqref{eq:H} yields Hamiltonian~\eqref{eq:HJn} with :
\begin{equation*}
J_n=-\mathcal{E}_J e^{-\phizpf^2/2}(-1)^n\left({\phizpf^{n}}/{n!}\right)^2\;.
\end{equation*}
{For completeness, we use the expression above to derive the expression of $J_n$ from Eq.\ \eqref{eq:Hcircuit}:
\begin{eqnarray}
J_n(\varphi_\text{ext})&=&-E_{J1}\cos(\varphi_\text{ext}) e^{-\varphizpf^2/2}(-1)^n\left({\varphizpf^{n}}/{n!}\right)^2\label{eq:Jngen}\\
&+&E_{J2}\cos(2\varphi_\text{ext}) e^{-(2\varphizpf)^2/2}(-1)^n\left({(2\varphizpf)^{n}}/{n!}\right)^2.\notag
\end{eqnarray}
}
We now turn to the derivation of Eq.\ \eqref{eq:deltanJn}. The eigenvectors of Hamiltonian~\eqref{eq:HJn} are Fock states $\ket{n}$ where $n$ is an integer. Using the formula: for $n\ge0$, $\ba^\dag\ket{n}=\sqrt{n+1}\ket{n+1}$ and for $n\ge 1$, $\ba\ket{n}=\sqrt{n}\ket{n-1}$ and $\ba\ket{0}=0$,  their associated eigenvalue $E_n$ take the form 
\begin{equation*}
E_n=n\hbar\omega_0+\sum_{k=2}^n\frac{n!}{(n-k)!}J_k\;.
\end{equation*}
We define the transition frequency $\omega_n=(E_{n+1}-E_n)/\hbar$, and $\delta_n=\omega_n-\omega_0$. Note that 
\begin{equation*}
E_{n+1}=(n+1)\hbar\omega_0+(n+1)!J_{n+1}+\sum_{k=2}^{n} \frac{(n+1)!}{(n+1-k)!}J_k\;,
\end{equation*}
so that 
\begin{align*}
\hbar\delta_n&=(n+1)!J_{n+1}+\sum_{k=2}^{n}  \left(\frac{(n+1)!}{(n+1-k)!}-\frac{n!}{(n-k)!}\right)J_k\\
&=(n+1)!J_{n+1}+n!\sum_{k=2}^{n} k \frac{n!}{(n+1-k)!}J_k\\
&=\sum_{k=2}^{n+1}  k\frac{n!}{(n+1-k)!}J_k\;.
\end{align*}


\section{Harmonics from single arm KITE}

\begin{figure}[t]
\centering
\includegraphics{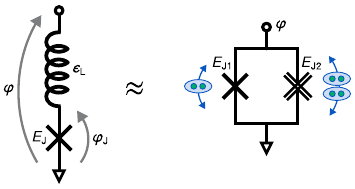}
\caption{{Electrical circuit diagrams depicting the emergence of Josephson harmonics when a linear inductance is connected in series with a Josephson junction without its self-capacitance (left). Imposing Kirchhoff's current law on the central node yields an effective potential containing both one- and two-Cooper-pair tunneling terms (right).}}
\label{fig:kepler}
\end{figure}

In this section we demonstrate that a Josephson junction in series with a small inductance generates Josephson harmonics \cite{Willsch2024,rymarz2023}. We carry out this simple derivation at zero frequency where capacitors may be disregarded. We consider the circuit depicted in Fig.\ \ref{fig:kepler}, in the regime where the inductive energy $\epsilon_L$ largely exceeds the Josephson energy $E_J$. We denote $\nu = E_J/\epsilon_L$ (here $\nu\ll 1$), $\varphi$ the phase drop across the entire circuit, and $\varphi_J$ the phase drop across the junction alone. The potential energy $U$ of this circuit is the sum of the inductive energy $U_L = \frac{1}{2}\epsilon_L(\varphi-\varphi_J)^2$ and the Josephson energy $U_J=-E_J\cos\varphi_J$. This yields $U = \epsilon_L\left[ \frac{1}{2} (\varphi-\varphi_J)^2 - \nu\cos\varphi_J\right]$. Also, $\varphi_J$ is tied to $\varphi$ though Kirchhoff's current law $E_J\sin\varphi_J=\epsilon_L(\varphi-\varphi_J)$. This equation may be recast as $\varphi_J=\varphi-\nu\sin\varphi_J$, and to first order in $\nu$, simplified to $\varphi_J=\varphi-\nu\sin\varphi+\mathcal{O}(\nu^2)$. Injecting this expression in the inductive energy yields $U_L = \frac{1}{2} \epsilon_L\left[\nu^2 \sin^2\varphi + \mathcal{O}(\nu^3)\right]$ and $U_J = -\epsilon_L\left[\nu\cos\varphi+\nu^2\sin^2\varphi + \mathcal{O}(\nu^3)\right]$. Summing these two energies results in $U = -\epsilon_L \left[\nu\cos \varphi + \frac{1}{2} \nu^2\sin^2\varphi + \mathcal{O}(\nu^3)\right]$. Finally, disregarding an irrelevant constant, we recover
\begin{equation}
U(\varphi) = -E_J\cos\varphi + \frac{1}{2}\frac{E_J^2}{2\epsilon_L}\cos2\varphi + \epsilon_L\mathcal{O}\left[(E_J/\epsilon_L)^3\right] \; .
\end{equation}
The first term in this expression is the usual Josephson energy that allows Cooper-pair tunneling. The second term is the second Josephson harmonic, and is responsible for the tunneling of pairs of Cooper pairs (Fig.\ \ref{fig:kepler}). The remaining perturbative term encompasses higher order harmonics.

\section{Model reduction}

\subsection{Outline of analysis}

{Separation of energy scales plays a central role in the understanding of complex systems. Remarkably, it is often possible to describe the dynamics of complex systems with just a handful of relevant low energy degrees of freedom. This concept has been pivotal for the description of molecules, and its most celebrated procedure is known as the Born-Oppenheimer approximation. Based on the fact that nuclei are much heavier than electrons, one computes the electron orbitals as a function of the nuclei positions. These orbitals then serve as the potential of the nuclei positional degree of freedom, accurately describing the quantized low-energy molecular vibrations.}

\begin{figure}[h]
\centering
\includegraphics{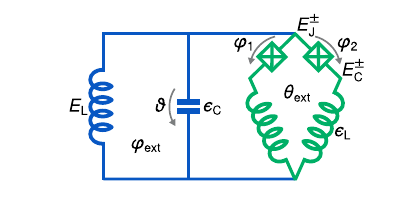}
\caption{Electrical circuit diagram of the $LC$ oscillator (blue) shunted by the KITE (green). As in the main text, we have $E_J^\pm = (1 \pm \varepsilon) E_J$ and $E_C^\pm = E_C / (1 \pm \varepsilon)$.} \label{fig:circuit}
\end{figure}

\begin{figure}[h]
\centering
\includegraphics{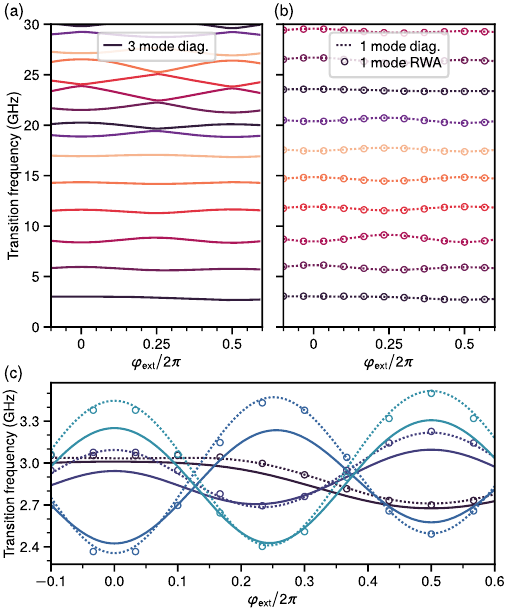}
\caption{Comparison of models of the device. (a--b) Transition frequencies from the ground state obtained from numerical diagonalization of the three-mode Hamiltonian in Eq.\ \eqref{eq:3mode_ham} at $\theta_\text{ext}=\pi$, as well as the one-mode Hamiltonian in Eq.\ \eqref{eq:1mode_ham_eval_simple}, and the RWA Hamiltonian of Eq.\eqref{eq:HJn} {with the expression of Eq.\ \eqref{eq:Jngen}}, both with the parameters of Tab.\ \ref{tab:params}. (c) Adjacent transition frequencies emphasizing the differences between the three models. \label{fig:model_reduction}}
\end{figure}

{In this section, we adapt this procedure to reduce the three-mode Hamiltonian of an $LC$ oscillator shunted by a KITE {(Fig.~\ref{fig:circuit})} to an effective one-mode Hamiltonian.} The procedure that we use is not mathematically rigorous, yet it provides a reduced model that agrees both qualitatively and, to a lesser extent, quantitatively with the full circuit model (see Fig.\ \ref{fig:model_reduction}). Improving the agreement and establishing a firm mathematical foundation will be the subject of future work \cite{Teufel2003}. We neglect the readout {and parasitic} resonators and consider the circuit in Fig.\ \ref{fig:circuit}. To realize the Hamiltonian in Eq.\ \eqref{eq:Hcircuit} at $\theta_\mathrm{ext} = \pi$, we first place ourselves in the parameter regime $E_L \ll \epsilon_C \sim E_C$, which guarantees large phase fluctuations in the oscillator. Second, we require $E_J \lesssim \epsilon_L \ll E_J / \varepsilon$ to lower the two-Cooper-pair tunneling energy below the bare Josephson energy $E_J$ while keeping single-Cooper-pair tunneling a perturbation. Finally, making the oscillator frequency the largest energy scale in the effective Hamiltonian (but still much smaller than the junction plasma frequency), we arrive at the parameter regime $E_L \ll \epsilon_C \sim E_C, E_J \lesssim \epsilon_L$ and $\varepsilon \ll 1$.
.

Our analysis is organized as follows. We start from the exact Hamiltonian for the circuit in Fig.\ \ref{fig:circuit} and move to a basis where the kinetic energy is diagonal and the modes are weakly coupled. In this basis, the Hamiltonian involves two high-frequency modes and one low-frequency mode. In the spirit of the Born-Oppenheimer approximation, we then assume that the high-frequency variables instantaneously minimize the potential energy for every value of the low-frequency variable. This yields an effective Hamiltonian for the low-frequency mode alone.

\subsection{Three-mode Hamiltonian and basis transformation}

We start with the circuit Hamiltonian \cite{Smith2022}
\begin{align}
\hat{H} &= \frac{2 E_C}{1 - \varepsilon^2} (\hat{N}_\Sigma^2 + \hat{N}_\Delta^2 - 2 \varepsilon \hat{N}_\Sigma \hat{N}_\Delta) + 4 \epsilon_C \hat{M}^2 + \frac{1}{2} E_L \hat{\vartheta}^2 \nonumber \\
&+ \epsilon_L \left(\hat{\vartheta} - \hat{\varphi}_\Sigma - \varphi_\mathrm{ext} - \tfrac{1}{2} \theta_\mathrm{ext} \right)^2 + \epsilon_L \left(\hat{\varphi}_\Delta - \tfrac{1}{2} \theta_\mathrm{ext}\right)^2 \nonumber \\
& - 2E_J \cos \hat{\varphi}_\Sigma \cos \hat{\varphi}_\Delta + 2\varepsilon E_J \sin \hat{\varphi}_\Sigma \sin \hat{\varphi}_\Delta \;, \label{eq:3mode_ham}
\end{align}
where we have introduced symmetric and antisymmetric variables $\hat{\varphi}_\Sigma = \frac{1}{2}(\hat{\varphi}_1 + \hat{\varphi}_2)$ and $\hat{\varphi}_\Delta = \frac{1}{2}(\hat{\varphi}_1 - \hat{\varphi}_2)$. Note that $\hat{\bm{N}} = \begin{pmatrix} \hat{N}_\Sigma & \hat{N}_\Delta & \hat{M}\end{pmatrix}^\mathrm{T}$ are the conjugate Cooper-pair numbers to $\hat{\bm{\phi}} = \begin{pmatrix} \hat{\varphi}_\Sigma & \hat{\varphi}_\Delta & \hat{\vartheta}\end{pmatrix}^\mathrm{T}$. Unfortunately, model reduction using the Born-Oppenheimer approximation would be inefficient in this basis due to the strong coupling between $\hat{\vartheta}$ and $\hat{\varphi}_\Sigma$ arising from the large value of $\epsilon_L$ relative to the other energy scales. 

To remedy this, we move to a basis where the dominant terms in the Hamiltonian are diagonal. The Hamiltonian in Eq.\ \eqref{eq:3mode_ham} may be decomposed into two parts: a linear Hamiltonian that accounts for the energy of the capacitances and internal KITE inductance (proportional to $E_C$, $\epsilon_C$, and $\epsilon_L$), and a nonlinear potential (proportional to $E_L$ and $E_J$). In our parameter regime, we can regard the latter as a perturbation and proceed by diagonalizing the former, which has the form
\begin{align*}
&\lim_{\substack{E_J \rightarrow 0 \\ E_L \rightarrow 0}}\hat{H} = 4 \hat{\bm{N}}^\mathrm{T} \mathbb{E}_C \hat{\bm{N}} + \tfrac{1}{2} \hat{\bm{\phi}}^\mathrm{T} \mathbb{E}_L \hat{\bm{\phi}} \\
&\mathbb{E}_C = \epsilon_C \begin{pmatrix} z & -\varepsilon z & 0 \\ -\varepsilon z & z & 0 \\ 0 & 0 & 1 \end{pmatrix} \qquad \mathbb{E}_L = 2 \epsilon_L \begin{pmatrix} 1 & 0 & -1 \\ 0 & 1 & 0 \\ -1 & 0 & 1 \end{pmatrix}
\end{align*}
upon a gauge transformation to shift $\hat{\varphi}_\Sigma \rightarrow \hat{\varphi}_\Sigma - \varphi_\mathrm{ext} - \frac{1}{2} \theta_\mathrm{ext}$ and $\hat{\varphi}_\Delta \rightarrow \hat{\varphi}_\Delta + \frac{1}{2} \theta_\mathrm{ext}$. We have also defined $z = \frac{E_C}{2 \epsilon_C (1 - \varepsilon^2)} = \frac{E_C}{2 \epsilon_C} + \mathcal{O}(\varepsilon^2)$ to parameterize the degree of capacitive loading, which diverges in the limit of vanishing junction capacitance. This linear Hamiltonian can be diagonalized with the transformation
\begin{align*}
&\hat{\bm{\phi}} = \Lambda \hat{\bm{\varphi}} \\
&\Lambda = \begin{pmatrix} 1 & (z-1)\varepsilon & -z \\ 0 & 1 & z(1+z)\varepsilon \\ 1 & -\varepsilon & 1 \end{pmatrix} + \mathcal{O}(\varepsilon^2) \; ,
\end{align*}
which has the effect of shuttling the coupling from terms of order $\epsilon_L$ to terms of smaller order $E_J$ and $E_L$. We denote the transformed superconducting phases $\hat{\bm{\varphi}} = \begin{pmatrix} \hat{\varphi} & \hat{\theta} & \hat{\zeta} \end{pmatrix}^\mathrm{T}$. Note that in the limit of vanishing junction asymmetry and capacitance, $z$ diverges and $\hat{\vartheta} \rightarrow \hat{\varphi}$. Recalling that the charging and inductive energy matrices transform according to $\mathbb{E}_C \rightarrow \Lambda^{-1} \mathbb{E}_C (\Lambda^\mathrm{T})^{-1}$ and $\mathbb{E}_L \rightarrow \Lambda^\mathrm{T} \mathbb{E}_L \Lambda$, we arrive at the Hamiltonian
\begin{align}
\hat{H} &= \frac{4 \epsilon_C}{1 + z^{-1}} \hat{N}_\varphi^2 + 2 E_C \hat{N}_\theta^2 + \frac{4\epsilon_C}{1 + z} \hat{N}_\zeta^2 \nonumber \\
&+ \epsilon_L \hat{\theta}^2 + \epsilon_L (1 + z)^2 \hat{\zeta}^2 \nonumber \\
&+ \frac{1}{2} E_L \left(\hat{\varphi} + \hat{\zeta} - \varepsilon \hat{\theta} + \varphi_\mathrm{ext} + \tfrac{1}{2} \theta_\mathrm{ext}\right)^2 \nonumber \\
&- 2 E_J \cos \left[ \hat{\varphi} - z \hat{\zeta} + (z-1) \varepsilon \hat{\theta} \right] \nonumber \\
&\qquad \times \cos \left[ \hat{\theta} + z (1+z)\varepsilon \hat{\zeta} + \tfrac{1}{2} \theta_\mathrm{ext} \right] \nonumber \\
&+ 2 \varepsilon E_J \sin \left( \hat{\varphi} - z \hat{\zeta} \right) \sin \left( \hat{\theta} + \tfrac{1}{2} \theta_\mathrm{ext} \right) + \epsilon_L \, \mathcal{O}(\varepsilon^2) \; , \label{eq:transformed_3mode_ham}
\end{align}
where we have introduced the transformed conjugate Cooper-pair numbers $\hat{N}_\varphi$, $\hat{N}_\theta$, and $\hat{N}_\zeta$; and shifted $\hat{\varphi} \rightarrow \hat{\varphi} + \varphi_\mathrm{ext} + \frac{1}{2} \theta_\mathrm{ext}$. Note that the charging energy of the $\hat{\varphi}$ mode approaches $\epsilon_C$ when the junction capacitance vanishes. On the other hand, when the shunting capacitance and junction asymmetry vanish, the Josephson potential terms in Eq.\ \eqref{eq:transformed_3mode_ham} are identical to those in Eq.\ \eqref{eq:3mode_ham}.

\subsection{Born-Oppenheimer approximation}

In our parameter regime, $\epsilon_L$ is the dominant energy scale. Therefore, the Hamiltonian in Eq.\ \eqref{eq:transformed_3mode_ham} describes two high-frequency modes $\hat{\theta}$ and $\hat{\zeta}$, and one low-frequency mode $\hat{\varphi}$, which are all perturbatively coupled through the terms proportional to $E_J$ and $E_L$. Our goal is to derive an effective Hamiltonian for the low-frequency $\hat{\varphi}$ mode. A method that is well suited for this purpose is the Born-Oppenheimer approximation, where one sets $\hat{\varphi}$ to a classical parameter $\varphi$ in Eq.\ \eqref{eq:transformed_3mode_ham} and removes the kinetic term proportional to $\hat{N}_\varphi^2$, yielding a Hamiltonian denoted $\hat{H}_{\theta\zeta}$. One then needs to calculate the ground state energy $E_0(\varphi)$ of $\hat{H}_{\theta\zeta}$. The Born-Oppenheimer approximation states that the effective potential of the $\hat{\varphi}$ mode is well captured by $\hat E_0(\hat{\varphi})$. However, an analytic expression for $E_0$ is unknown. Instead, we place ourselves in a semi-classical limit where the fluctuations of $\hat\theta$ and $\hat\zeta$ are neglected, and we make the approximation $\hat E_0(\hat\varphi)\approx \hat{U} \Big|_{\hat{\theta}_0, \hat{\zeta}_0}$, where $\hat U$ is the potential energy in Eq.\ \eqref{eq:transformed_3mode_ham} that we evaluate at the values $\hat{\theta}_0$ and $\hat{\zeta}_0$ that minimize $\hat U$ for every value of $\hat{\varphi}$.

We recall that we are placed in the parameter regime where $E_L \ll E_J \lesssim \epsilon_L$. In the following, we will assume $E_J / \epsilon_L = \mathcal{O}(\lambda)$ and $E_L / \epsilon_L = \mathcal{O}(\lambda^2)$, where $\lambda$ is a small parameter. The equilibrium values themselves are found by solving $\frac{\partial \hat{U}}{\partial \hat{\theta}}\big|_{\hat{\theta}_0, \hat{\zeta}_0} = 0$ and $\frac{\partial \hat{U}}{\partial \hat{\zeta}}\big|_{\hat{\theta}_0, \hat{\zeta}_0} = 0$. Retaining only the lowest-order terms in $\lambda$ and $\varepsilon$, we find
\begin{align}
\hat{\theta}_0 &= - \frac{E_J}{\epsilon_L} \left[ \sin \tfrac{1}{2} \theta_\mathrm{ext} \cos \hat{\varphi} + \mathcal{O}(\lambda, \varepsilon) \right] \nonumber \\
\hat{\zeta}_0 &= \frac{E_J z}{\epsilon_L (1 + z)^2} \left[ \cos \tfrac{1}{2} \theta_\mathrm{ext} \sin \hat{\varphi} + \mathcal{O}(\lambda, \varepsilon) \right] \; . \label{eq:equilibrium}
\end{align}

The low-frequency Hamiltonian can then be calculated---in the semiclassical limit of vanishing fluctuations of the phases $\hat{\theta}$ and $\hat{\zeta}$---to be
\begin{align}
\hat{H}_\varphi &= \frac{4 \epsilon_C}{1 + z^{-1}} \hat{N}_\varphi^2 + \hat{U} \Big|_{\hat{\theta}_0, \hat{\zeta}_0} \nonumber \\
&= \frac{4 \epsilon_C}{1 + z^{-1}} \hat{N}_\varphi^2 + \frac{1}{2} E_L \left(\hat{\varphi} + \varphi_\mathrm{ext} + \tfrac{1}{2} \theta_\mathrm{ext}\right)^2 \nonumber \\
&- \frac{E_J^2}{2 \epsilon_L} \left[ \sin^2 \tfrac{1}{2} \theta_\mathrm{ext} - \frac{z^2}{(1+z)^2} \cos^2 \tfrac{1}{2} \theta_\mathrm{ext} \right] \cos 2 \hat{\varphi} \nonumber \\
&- 2 E_J \cos \tfrac{1}{2} \theta_\mathrm{ext} \cos \hat{\varphi} + 2\varepsilon E_J \sin \tfrac{1}{2} \theta_\mathrm{ext} \sin \hat{\varphi} \nonumber \\
&+ \lambda E_J \, \mathcal{O} (\lambda, \varepsilon) + E_C \, \mathcal{O}( \varepsilon^2) \; . \label{eq:1mode_ham_simple}
\end{align}
Finally, shifting $\hat{\varphi} \rightarrow \hat{\varphi} - \varphi_\mathrm{ext} - \frac{1}{2} \theta_\mathrm{ext}$ and setting $\theta_\mathrm{ext} = \pi$ yields the single-mode model referenced in the main text,
\begin{align}
\hat{H}_\varphi \Big|_\pi &= \frac{4 \epsilon_C}{1 + z^{-1}} \hat{N}_\varphi^2 + \frac{1}{2} E_L \hat{\varphi}^2 + \frac{E_J^2}{2 \epsilon_L} \cos \left[2(\hat{\varphi} - \varphi_\mathrm{ext})\right] \nonumber \\
&- 2 \varepsilon E_J \cos (\hat{\varphi} - \varphi_\mathrm{ext}) \nonumber \\
&+ \lambda E_J \, \mathcal{O}(\lambda, \varepsilon) + E_C \, \mathcal{O} (\varepsilon^2) 
\label{eq:1mode_ham_eval_simple}
\end{align}
after neglecting constant terms. We see that the shunting capacitance renormalizes the charging energy, while the inductive energy is completely unchanged. The first, second, and third terms in the potential represent the contributions of the inductive shunt, two-Cooper-pair tunneling, and asymmetric dc-SQUID, respectively.

The Hamiltonian in Eq. \eqref{eq:1mode_ham_eval_simple} is the main result of this section, and it explains both the approximate values of the effective parameters $\Omega$, {$E_{J1}$, $E_{J2}$, and $\varphizpf$} of the low-energy Hamiltonian as well as the functional form. Overall, we sought a lowest-order result in the parameters $\varepsilon$ and $\lambda$, and completely neglected fluctuations of the high frequency modes. The parameters achieved in the experiment are not especially deep in this parameter regime---for instance $E_J / \epsilon_L = 0.58$ and the fluctuations of $\hat{\theta}$ are of order $\left(\frac{E_C}{2\epsilon_L}\right)^{1/4} = 0.82$ (similarly for $\hat{\zeta}$)---making the effective parameters in Eq.\ \eqref{eq:1mode_ham_eval_simple} differ slightly from those found using a fit (see Tab.\ \ref{tab:params2}). Improved accuracy is likely possible using higher order corrections in $\lambda$ and including fluctuations.


\section{Full lumped element model}
\begin{figure}[h]
\centering
\includegraphics[width=0.5\textwidth]{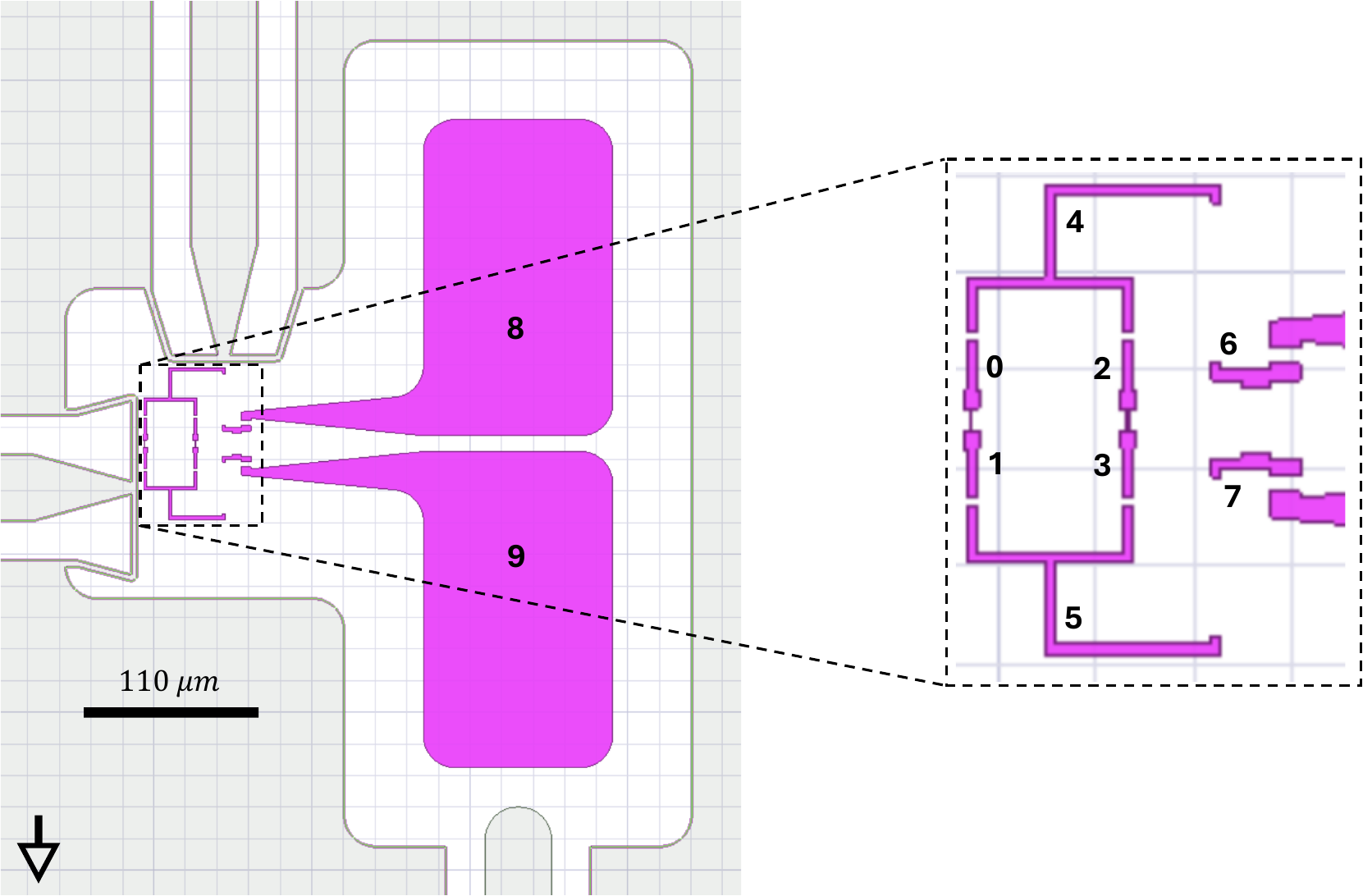}
\caption{{Layout of the circuit chip simulated in ANSYS Q3D in order to extract the capacitance matrix.} \label{fig:layout}}
\end{figure}

\begin{figure}[h]
\centering
\includegraphics[width=0.5\textwidth]{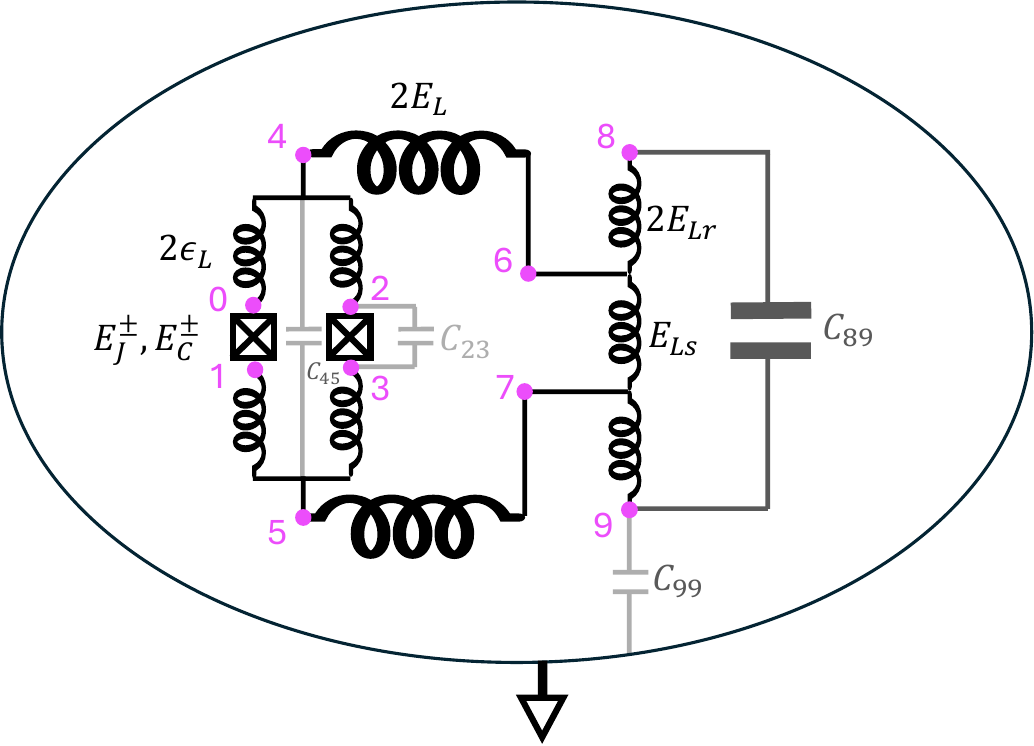}
\caption{{Lumped element circuit of our sample. Pink nodes have an all-to-all capacitive coupling. The corresponding capacitance matrix is displayed in Table.\ \ref{tab:capacitance}. Only some capacitances are represented for clarity. \label{fig:circuit_full}}}
\end{figure}

{In this section we describe the procedure we follow in order to fit the data of Fig.\ \ref{fig:fig4}.
\paragraph{Capacitance matrix: }
We start by exporting the layout file of our circuit to the finite element solver {Ansys} Q3D (see Fig.\ \ref{fig:layout}). We then remove all Josephson junctions (JJs). This involves both the two small KITE junctions, as well as all the JJ arrays that form the various lumped inductances of our circuit. At this stage our layout contains 10 floating electrodes (labeled 0 to 9 in Fig.\ \ref{fig:layout}), as well as a ground plane and an input line that are toggled to ground. The software then returns the capacitance matrix (see Tab.\ \ref{tab:capacitance}). In order to build the full capacitance matrix, we add the capacitive contribution of the two small KITE junctions. A further refinement of our model{---}not implemented here{---}would be to add the capacitive contribution of the JJ arrays forming the inductors
\paragraph{Inductance matrix and Josephson energy: }
We then proceed to constructing the inverse inductance matrix from the inductive energy of the JJ arrays. Each array is simply replaced by a linear inductor. A further refinement of our model{---}not implemented here{---}would be to account for the nonlinearity of the JJ arrays, as well as array modes. Finally, we include the Josephson cosine potential associated to the two small KITE junctions. At this stage we have obtained the full Lagrangian of the lumped element circuit depicted in Fig.\ \ref{fig:circuit_full}.
\paragraph{Hamiltonian:} We then obtain the Hamiltonian by following the standard circuit quantization procedure (see e.g. Ref.\ \cite{Smith_thesis_2019}). We only retain the {five} lowest frequency modes, that is: the three KITE modes, the parasitic, and the readout modes.
\paragraph{Fitting procedure: } Our approach is to assume that the geometry of our circuit is well known, and hence the capacitance matrix extracted from {Ansys} Q3D should be fixed. On the other hand, the inductances corresponding to the JJ arrays, as well as the Josephson and charging energies of the two small KITE junctions are not precisely known. Indeed, these quantities are sensitive to nanofabrication parameters (oxidation pressure and duration). In principle, Josephson inductances could be extracted from room temperature measurements through the Ambegaokar-Baratoff formula. However, this formula contains the superconducting gap of the junction electrodes made of thin film aluminium, as well as the value of the resistance right above the critical temperature, which are difficult to access. Finally, Josephson tunnel barriers are known to vary with time through aging. Consequently, we use as our fit parameters: $E_J, E_C, \epsilon_L, E_L, E_{LR},E_{LS}$, and the asymmetry parameter $\varepsilon$. The results of the fitting procedure are shown in Tab.\ \ref{tab:params} of the main text.
\paragraph{Methodology: } Diagonalizing a Hamiltonian of {five} bosonic modes is a technical challenge. A direct approach would consist in truncating each mode to a maximal number of excitations (e.g. {$35$, $9$, $4$, $4$, $4$} for the three KITE modes, the readout mode, and the parasitic mode respectively). We would then need to diagonalize a Hamiltonian matrix whose storage size is {$\SI{300}{\mega\byte}$}. Diagonalizing such a large matrix for hundreds of external flux values, and along a gradient descent fitting procedure would be prohibitively time consuming.}
{Instead we perform a hierarchical diagonalization. The full {five-}mode Hamiltonian may be decomposed as the sum of three terms: one only involving the three KITE modes, one only involving the readout and parasitic modes, and a coupling term involving all {five} modes. We first diagonalize the KITE modes Hamiltonian {[size $(35\times9\times4)^2$]} and then retain only the $20$ lowest states. We then propagate this change of frame and projection on the coupling Hamiltonian. We are left with a matrix of size {$(20\times4\times4)^2$}. The matrix storage size is {$\SI{30}{\kilo\byte}$}, which is a $10^4$ reduction with respect to the direct method described above.}

\begin{table}[h]
    \centering
    {
        \begin{tabular}{c|cccccccccc}
        \toprule
         & 0  & 1  & 2  & 3  & 4  & 5  & 6  & 7  & 8  & 9  \\
        \hline
         0& 2.13 & -0.21 & -0.04 & -0.03 & -0.34 & -0.08 & -0.01 & -0.01 & -0.04 & -0.03 \\
        1&& 2.13 & -0.03 & -0.04 & -0.07 & -0.34 & -0.01 & -0.01 & -0.03 & -0.04 \\
        2&& & 1.97 & -0.28 & -0.45 & -0.12 & -0.11 & -0.07 & -0.12 & -0.1 \\
        3&& & & 1.97 & -0.11 & -0.47 & -0.07 & -0.11 & -0.1 & -0.13 \\
        4&& & & & 8.86 & -0.19 & -0.18 & -0.08 & -0.5 & -0.27 \\
        5&& & & & & 7.98 & -0.09 & -0.21 & -0.33 & -0.73 \\
        6&& & & & & & 2.6 & -0.24 & -0.87 & -0.29 \\
        7&& & & & & & & 2.6 & -0.28 & -0.89 \\
        8&& & & & & & & & 69.23 & -18.73 \\
        9&& & & & & & & & & 69.39 \\
\toprule
    \end{tabular}
    }
    \caption{{Capacitance matrix in femtofarads. Rows and columns correspond to the metallic plates labeled in Fig.\ \ref{fig:layout}. The matrix is symmetric and the lower diagonal is omitted for clarity. The entries $m_{ij}$ of this matrix are related to the capacitance $C_{ij}$ between plates $i$ and $j$ as follows: $C_{ij}=-m_{ij}$ for $i\ne j$ and the capacitance to ground of element $i$ is $C_{ii}=\sum_j{m_{ij}}$.}}
    \label{tab:capacitance}
\end{table}

\section{Extended data}

\begin{figure}[h]
\centering
\includegraphics{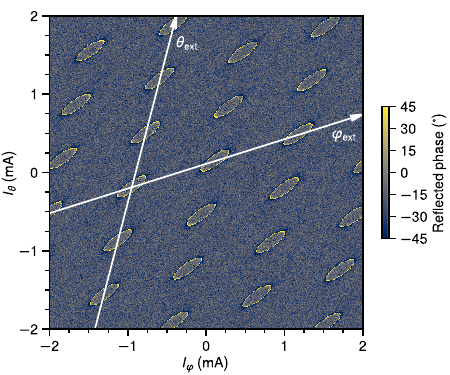}
\caption{Phase of a reflected signal at frequency $\SI{5.05532}{\giga\hertz}$ on the readout resonator as a function of the two current sources depicted in Fig.\ \ref{fig:device}. The phase response is periodic in two directions, which we identify with white arrows to be the independent external fluxes $\varphi_\mathrm{ext}$ and $\theta_\mathrm{ext}$. \label{fig:flux_map}}
\end{figure}

\paragraph{Flux calibration: }
Due to their layout, our physical flux bias lines (see Fig.\ \ref{fig:device}) inevitably couple to both circuit external fluxes $\varphi_\mathrm{ext}$ and $\theta_\mathrm{ext}$. For initial characterization, we probe the readout resonator at the single frequency $\SI{5.05532}{\giga\hertz}$ and measure the reflected phase as a function of both bias currents. As the resulting pattern is 2D-periodic with many flux quanta visible, we can easily find the affine transformation that maps bias currents to $\varphi_\mathrm{ext}$ and $\theta_\mathrm{ext}$, as shown in Fig.\ \ref{fig:flux_map}. Additionally, we observe external flux drifts on the order of $1$--$\SI{2}{\percent}$ of a flux quantum on daily timescales, which we calibrate regularly using the procedure described in Ref.\ \citealp{Smith2022}.

\paragraph{{Zoom on spectroscopy sweeps:}}
{It is difficult to resolve our circuit's sharp resonance features on the wide span of Fig.\ \subref*{fig:twotone_varphi_half}. Therefore, in Fig.\ \ref{fig:spectroscopy_zoom_cuts}, we zoom on this dataset. The resonances are easily found and marked with open circles. In order to average out small flux drifts during the data acquisition, these spectroscopy sweeps are acquired multiple times. The corresponding resonances are averaged together to yield the data points of Figs.\ \ref{fig:fig5}--\ref{fig:fig6}.}

\begin{figure*}[h]
\centering
\includegraphics{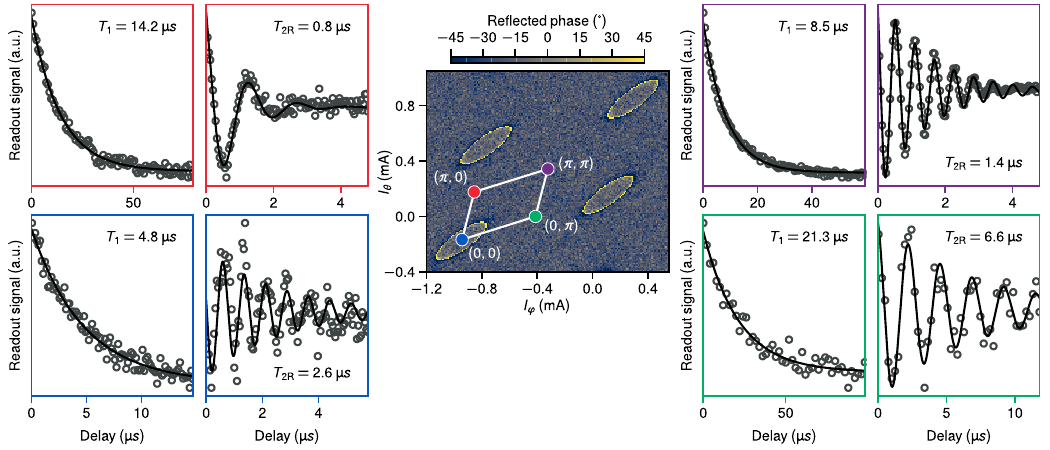}
\caption{Qubit coherence times measured at the four distinct flux sweet spots, as indicated by colored circles in the central panel, which is a zoom of the wide external flux map in Fig.\ \ref{fig:flux_map}. The sweet spots are labeled by their coordinates $(\theta_\mathrm{ext}, \varphi_\mathrm{ext})$, and the white lines connecting them correspond to the axes used for the spectroscopy data in Fig.\ \ref{fig:fig4}. \label{fig:coherence}}
\end{figure*}

\begin{figure*}
    \centering
    \includegraphics[width=0.7\linewidth]{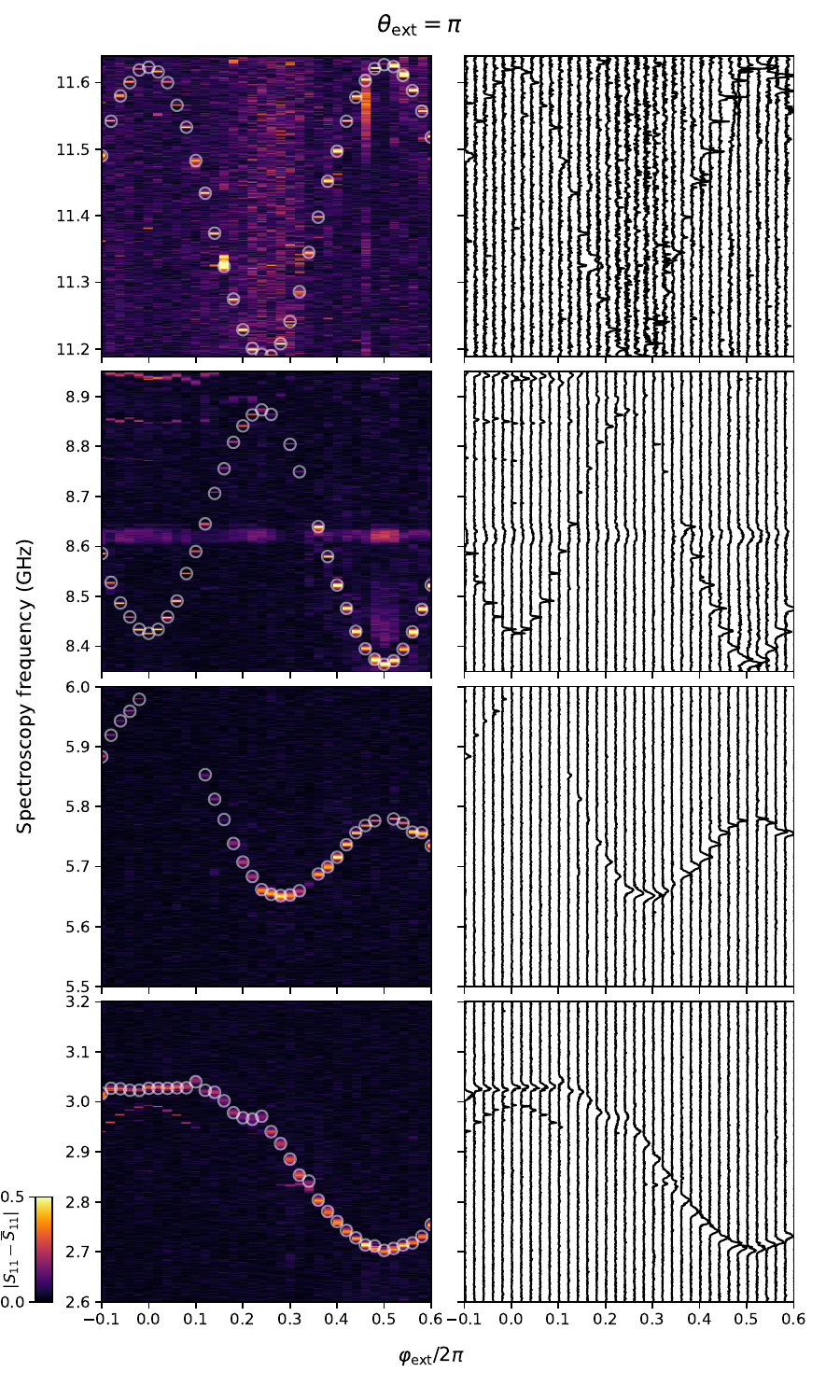}
    \caption{{Zoom on dataset of Fig.\ \protect\subref*{fig:twotone_varphi_half} around the $\ket{0}\rightarrow\ket{n}$ transitions ($n=1,2,3,4$). Resonances are marked with white open circles. Right column: the same dataset as the left column presented as one-dimensional cuts. }}
\label{fig:spectroscopy_zoom_cuts}
\end{figure*}

\paragraph{Coherence times :}
In addition to the spectroscopy data shown in the main text, we acquire coherence data of the lowest frequency transition---the qubit---at the four external flux sweet spots, as shown in Fig.\ \ref{fig:coherence}. As visible in Fig.\ \ref{fig:fig3}, the qubit frequency varies by an order of magnitude from $\SI{636}{\mega\hertz}$ to $\SI{6.882}{\giga\hertz}$ as the external flux point is stepped from $(\theta_\mathrm{ext}, \varphi_\mathrm{ext}) = (0, \pi)$ to $(\pi, \pi)$ to $(\pi, 0)$ to $(0, 0)$; in other words, counter-clockwise around the plaquette in Fig.\ \ref{fig:coherence} starting at the green point. As the qubit frequency increases, the observed relaxation times decrease from $\SI{21.3}{\micro\second}$ to $\SI{4.8}{\micro\second}$, indicating a roughly constant quality factor loss channel such as dielectric loss. The coherence times measured with a Ramsey sequence range from $0.8$--$\SI{6.6}{\micro\second}$ and improve by about a factor of two using a single echo pulse (not shown). The dephasing is likely due to a combination of second-order flux noise, phase-slips in the array junctions, and photon shot noise in the readout resonator.

\end{document}